# Cepstrum-based interferometric microscopy (CIM) for quantitative phase imaging

Ricardo Rubio-Oliver[a], Javier García[a], Zeev Zalevsky[b], José Ángel Picazo-Bueno[b, c], Vicente Micó[a, *]

[a] *Departamento de Óptica y de Optometría y Ciencias de la Visión. Facultad de Física. Universidad de Valencia. C/ Doctor Moliner 50, 46100 Burjassot, Spain*

[b] *Faculty of Engineering, Bar-Ilan University, Ramat-Gan 52900, Israel*

[c] *Biomedical Technology Center, University of Muenster, Mendelstr. 17, D-48149 Muenster, Germany*

*Email: vicente.mico@uv.es

**Abstract:** A universal methodology for coding-decoding the complex amplitude field of an imaged sample in coherent microscopy is presented, where no restrictions on any of the two interferometric beams are required. Thus, the imaging beam can be overlapped with, in general, any other complex amplitude distribution and, in particular, with a coherent and shifted version of itself considering two orthogonal directions. The complex field values are retrieved by a novel Cepstrum-based algorithm, named as Spatial-Shifting Cepstrum (SSC), based on a weighted subtraction of the Cepstrum transform in the cross-correlation term of the object field spectrum in addition with the generation of a complex pupil from the combination of the information retrieved from different holographic recordings (one in horizontal and one in vertical direction) where one of the interferometric beams is shifted 1 pixel. As a result, the field of view is tripled since the complex amplitudes of the three interferometric fields involved in the process are retrieved. Proof-of-concept validation of this methodology, named as Cepstrum-based Interferometric Microscopy (CIM), is provided considering an off-axis holographic configuration for retrieving the cross-correlation of the two interferometric complex amplitude fields in a compact quasi-common path Michelson interferometric configuration. Experimental results for different types of phase samples (resolution test targets for step-by-step calibration and demonstration as well as fixed biosamples) are included.

**Keywords:** digital holographic microscopy, quantitative phase imaging, phase retrieval, Cepstrum transform, Cepstrum-based interferometric microscopy.

## 1. Introduction

Quantitative phase imaging (QPI) is a modern and widely used discipline in optical microscopy for phase imaging of biological samples [1,2]. QPI is capable of accurately extracting quantitative information in the optical phase delay of light passing through a sample, caused by the topography and/or spatial refractive index distribution in the specimen. QPI allows the measurement of biophysical cell parameters for morphological studies of biological cells, physical parameters extraction, cell phenotyping and histopathology, even for long-term label-free observation of large cell populations [3–5]. Perhaps the most common way to achieve QPI is by using digital holographic microscopy (DHM) [6]. Implemented in a single imaging platform, DHM combines optical light microscopy, interferometric accuracy, digital imaging and computer-assisted data retrieval to provide whole-object wavefront recovery with numerical processing capabilities in a non-invasive (non-stained), full-field (non-scanning), non-contact (no sample damage) and static (no moving components) operating principle [7–9]. In addition, off-axis DHM provides a real-time (on-line control) capability based on single hologram acquisition at the cost of limiting the system requirements in terms of sensor space–bandwidth, that is, by generating an empty space in the spatial frequency domain to allow Fourier filtering process [10].

Mainly in the last decade, strong efforts have been placed to validate compact, miniaturized and simple experimental interferometric arrangements for QPI such as using a Lloyd's mirror [11], different types of prisms [12,13], diffraction gratings [14,15], beam splitter cubes [16,17] and Gabor-like schemes [18–20], just to cite a few. Moreover, it is extremely appealing to implement coherent sensing capabilities to commercially available bright field microscopes in order to convert them into QPI microscopes with minimum modifications and in a cost-effective way. Some examples have been reported by adapting external add-on modules based on wavefront sensing [21–23], lateral shearing interferometers [24–26], transport of intensity equation methodologies [27,28], with a purely numerical package [29], and using Mach-Zehnder [30], Sagnac [31,32], Michelson [33–36] as well as common-path [37,38] based layouts. The combination of these last approaches also yields in doubling [39,40] and tripling [41] the field of view (FOV) provided by the technique.

However, strong restrictive conditions must be applied to all of these QPI approaches for their correct implementation. Since holography is based on the interference between the target beam with a clear reference beam, some approaches [11,12,24–26,32,33,37,38] assume that the sample is confined to a specific limited area (sparse sample constraints) thus leaving a surrounded free space acting as reference beam. Others [13–17] directly preserve a clear reference region at the input plane (spatial multiplexing) to transmit both interferometric beams in parallel through the same microscope platform. And other approaches split the imaging beam in two beams while introducing a pinhole in one of them to filter out a clean reference beam [30,31,34–36], similar to diffraction phase microscopy [42] and some approaches in synthetic aperture microscopy [43,44]. Thus, methods based on sparse sample constraints limit their applicability to dense samples, approaches introducing spatial multiplexing limit the allowable FOV, and the pinhole requirement complicates the layout from a hardware point of view.

Here, we introduce Cepstrum-based Interferometric Microscopy (CIM) as a completely novel methodology for QPI in DHM solving the previously mentioned drawbacks, that is, no restrictions are introduced to any of the interferometric beams (no sample constraints nor any type of spatial multiplexing to allow clear reference beam transmission) and no complicated to align optical elements (pinholes) are needed making bulky systems. CIM is based on the coherent overlapping of two arbitrary complex fields which, in this contribution will be implemented in an off-axis configuration. Thus, there is no need to introduce a clear reference beam at the recording plane, as is usually done in DHM, because the imaging beam can be overlapped with any other complex amplitude field. The procedure requires two interferometric recordings, one in conventional off-axis mode and another off-axis hologram considering a single camera pixel shift in one of the interferometric images, for two orthogonal directions (here implemented as horizontal and vertical but not necessarily restricted to them), where one of the complex fields is constant. After applying a weighted Cepstrum subtraction operation of the cross-correlation term of the different off-axis holograms and the combination of the retrieved information using complementary bow-tie masks at the Fourier domain to synthesize the complex pupil of the static field QPI values are extracted from the four recorded digital holograms. Thus, the outcomes are the retrieval of the complex values for the 3 different images involved in the process meaning that QPI is accomplished while tripling the FOV. We have named this Cepstrum-based algorithm as Spatial-Shifting Cepstrum (SSC) as a spatial counterpart to the common phase-shifting approaches in on-axis DHM.

In fact, the two-step phase-shifting algorithm proposed by Zhang et al. for phase retrieval in on-axis holography retrieves phase information based on a logarithmic transformation of the recorded hologram Fourier transform [45]. This numerical manipulation implicitly involves a cepstrum transformation and can be understood as the first implementation by simulations of cepstrum algorithm applied to holography. Experimental validation of the Zhang et al. cepstrum methodology was recently proposed by Pan et al. and improved with single-shot operation principle by using two digital cameras [46]. Cepstrum has been also proposed in DHM for slightly-off-axis architectures [47], where they employ a similar approach to the one used by Zhang et al., yet requiring only a single hologram capture. However, all these algorithms work only in the case where the reference beam is much brighter than the object beam, and more importantly, they require a known clean reference, while our proposed SSC has none of these constraints.

To summarize, in this contribution we implement CIM proof-of-concept validation in a compact quasi common-path Michelson interferometric configuration assembled after the tube lens in a classical infinity corrected microscope configuration, where one of the mirrors is movable to change the off-axis direction as well as to allow the pixel shift required for the application of the SSC algorithm. Experimental results for different types of phase samples (resolution test targets and fixed biosamples) are presented validating the proposed CIM concept as a completely novel procedure for accurate QPI with tripled FOV in DHM.

## 2. Materials and methods

*2.1 Optical Layout and Principle of CIM*

The experimental arrangement for CIM implementation is presented in Fig. 1. It is based on a Michelson interferometric configuration assembled after the tube lens of a conventional infinity corrected microscope configuration. In the Michelson layout, one mirror (M1) is static while the other one (M2) is tilted to allow off-axis recording and mounted onto a linear translation stage to provide the necessary image shift for SSC algorithm application. The recording process of the different holograms (4 in total) needed to obtain the QPI of the 3 different FOVs is done sequentially in time.

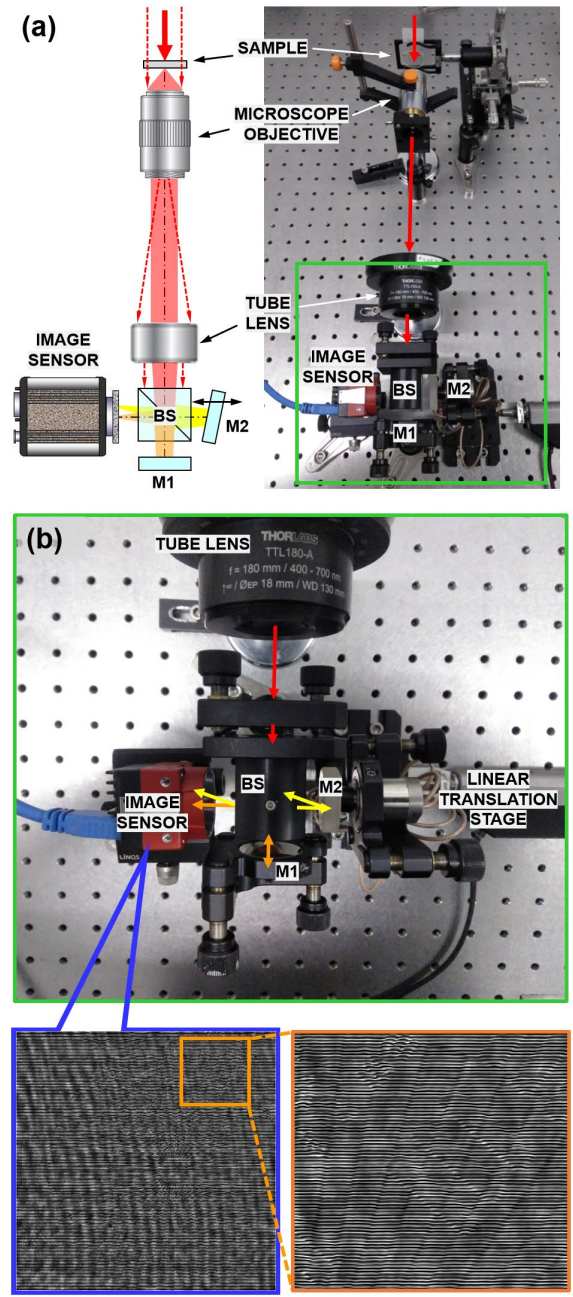

Fig. 1. Optical layout of CIM. (a) Scheme (left) and picture at the lab (right) for the implementation of CIM. (b) Magnification of the green rectangle in (a) showing the optical arrangement after the tube lens. Case (b) also includes one example of the recorded holograms (one hologram related with Fig. 2, step num. 7) showing the overlapping of two different FOVs (part of the USAF resolution test + part of the spoke target) in addition with an inset to clearly see the interferometric fringes. The main components have been identified for clarity. BS, beam splitter cube; M1, static mirror; M2, movable tilted mirror.

This lab-made imaging platform is composed by a microscope objective (Mitutoyo Infinity Corrected Long Working Distance Objective 7.5X 0.21NA) and an infinity corrected tube lens (Thorlabs TTL180-A, 180 mm focal length). After the tube lens, a compact (10 cm x 12 cm) Michelson interferometer consisting on a non-polarizing beam splitter cube (Thorlabs BS013, 50:50 R-T, 25.4mm side) and two round protected aluminium mirrors (Thorlabs PF10-03-G01, 25.4 mm diameter) is implemented. Due to the off-axis nature of the interferometric layout, all the components (BS, M1 and M2) are mounted into kinematic mounts (Thorlabs KM100) to effectively control the tilt angle between the interferometric beams. Finally, a digital imaging sensor (Allied Vision Alvium C 1800U-811m C USB3, 2.74 μm × 2.74 μm, 2848 x 2848 pixel resolution) connected to a standard laptop computer (Dell XPS15 9550) records and numerically processes (Matlab version 2021b) the holograms.

Illustration about how CIM performs is outlined through Fig. 2 considering the case of a phase resolution test target including USAF-like and focus star patterns having different heights (from 50 to 350 nm at 50 nm steps). Along the figure, the process is guided by numbers identifying the steps to be followed. Starting at the play symbol (►), the selected region of interest (ROI) is composed by patterns with 3 different heights (250, 300 and 350 nm) where the 3 multiplexed FOVs are color coded in blue/green/orange to easily track them during the process. Fig. 2 (up-right) also includes a wide FOV image obtained with a low magnification objective (Mitutoyo Plan Apo Infinity Corrected Long Objective, 2X, 0.055NA) to easily allocate the 3 multiplexed FOVs in the test slide. However, we have rotated the test target (step num. 1) in order to provide adequate test overlap between the different FOVs. This is not a critical fact but as a consequence of both the empty spaces in the test (test details are not presented in a continuous way through the test slide) and the limited flexibility in the lateral shift introduced by the off-axis recording (the shift required between each pair of images is dictated by the Fourier separation of the diffraction orders, since complex amplitude filtering is applied at the Fourier domain).

CIM implements a methodology based on the sequential acquisition of a total of 4 holograms, 2 in the horizontal and 2 in the vertical direction. Thus, the off-axis recording in the horizontal direction (step num. 2) allows the interferometric overlapping of the static image (blue FOV) with a shifted version of itself in the horizontal direction (orange FOV). This is our first hologram in the horizontal direction. The second one comes from the interference between the static image with the orange FOV considering an additional 1-pixel horizontal shift in the last one. The precise control of 1-pixel shift is accomplished by using the linear translation stage on the M2 mirror. Both holograms are under identical off-axis holographic configurations and the complex amplitude of the cross-correlation information coming from the two complex pupils can be retrieved by digital fast Fourier transform (FFT) and Fourier filtering of one diffraction order (step num. 4). These two complex images will be the horizontal inputs for the SSC algorithm. The same procedure is applied to the vertical direction (blue and green FOVs) by manually changing the off-axis angle in the M2 mirror mount (steps num. 3 and 5). Note that the angle formed between vertical and horizontal off-axis configurations is not critical, so that manual change of the off-axis tilt will not introduce any problem of accuracy to the SSC performance. Thus, the set of 4 off-axis holograms is obtained in time sequence in a similar way to, for instance, on-axis phase-shifting DHM implementations.

Each pair of 1-pixel shifted complex images is used to synthesize the complex spectrum of the static image (blue FOV). Further details are given in the SSC algorithm section. Qualitatively, the two horizontal complex fields can be used to retrieve the static image using the proposed SSC algorithm (step num. 6). However, there is a null direction providing useless values at the vertical central line of the FFT because of the mathematical SSC algorithm definition (see Eq. 5 at SSC algorithm section). This direction is the vertical one when considering the horizontal 1-pixel shift (and vice versa). This is the reason why there is a strong vertical line with infinite value in the spectrum retrieved by step num. 6 in Fig. 2. Similar procedure and findings (step num. 7) happen for the case of the two vertical complex images where now, the useless direction is provided in the horizontal (see the strong horizontal line at the center of the FFT in step num. 7). For this reason, we have implemented a final synthesis step of the static image spectrum (step num. 8) where each partially retrieved spectrum is multiplied by a complementary bow-tie mask in the Fourier domain, disregarding the useless direction. The final result retrieves the full spectrum corresponding to the static image except for an ambiguity at the spectrum center coming from the intersection of the two useless directions. We have called it as zero ambiguity since it is placed at the DC term of the retrieved spectrum. These zero ambiguities at the Fourier domain provides background inhomogeneities when performing inverse FFT in a similar way as the transport of intensity equation (TIE) algorithm does [48], but it is not critical since the relative phase information of the object is fully retrieved. High-pass filtering (step num. 9) homogenizes the background level allowing quantitative phase contrast visualization of the retrieved complex image (blue FOV). Finally, the proposed algorithm recovers the two additional FOVs used in the process (step num.10 and 11) by subtracting the complex amplitude recovered for the static image from one of the two complex fields for both horizontal (orange FOV) and vertical (green FOV) multiplexed directions. As final outcome, QPI of the 3 involved FOVs is accessible with similar image quality than the retrieved for the static field.

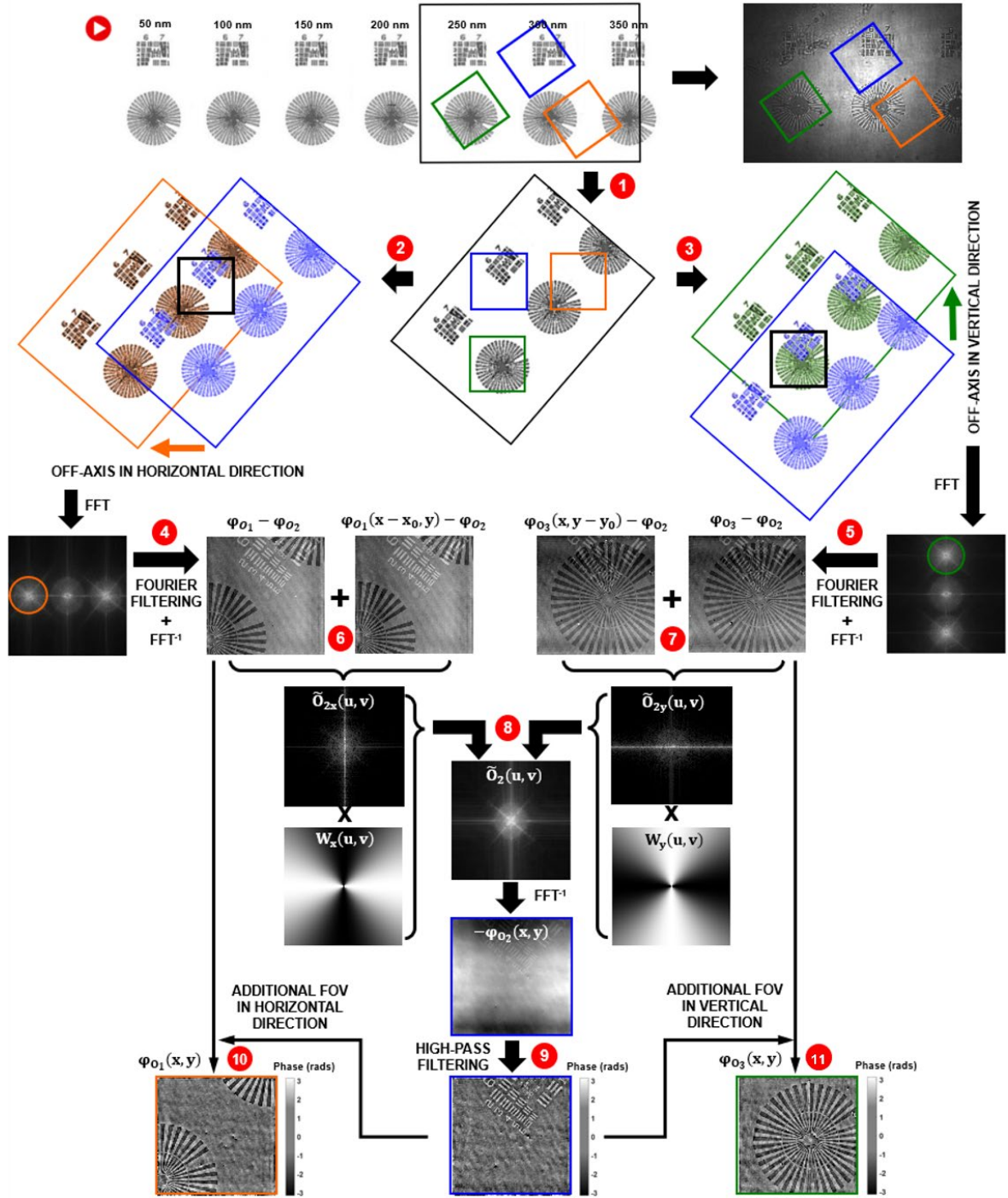

Fig. 2. Workflow of CIM. This hybrid scheme represents how CIM operates including a mix of drawings and experimental results to finally achieve QPI with a tripled FOV: starting from the identification of the 3 FOVs involved in the process (step 1) and following the different operations step by step (from 1 to 11 in red), CIM needs the recording of 4 off-axis holograms (2 in horizontal and 2 in vertical directions – steps 2 and 3) where one hologram

for each orthogonal direction comes from 1-pixel shift in one of the interferometric images. After filtering the complex amplitude distribution at the Fourier domain (steps 4 and 5), the 4 complex fields are combined by pairs using a SSC algorithm to retrieve the complex amplitude distribution of the static image (blue FOV) but containing a useless direction at Fourier domain (steps 6 and 7). The combination of both outcomes restricts to the DC term the ambiguity (step 8) which is removed by high-pass filtering process (step 9). Finally (steps 10 and 11), the additional FOVs in horizontal (orange) and vertical (green) directions are retrieved by subtraction from one of the complex distributions for every orthogonal direction. Notice that, from step 4, the images included in this figure are experimental results provided by CIM.

## 2.2 Calibration Procedure

The proposed methodology involves a precise shift in one of the interferometric images for every orthogonal direction being used. To accurately know and control such displacement (mirror M2 at Fig. 1), we have used a precision motor actuator (Newport LTA) running at a constant speed. Thus, for a given working frame rate and off-axis tilt, forty frames integrate a displacement of a single pixel in one of the interferometric images. The recording is done once the motor has reached the steady velocity and before deceleration begins. Under such conditions, we have estimated a confidence in the determination of the lateral displacement of about 25 thousandths of pixel. This procedure was calibrated using an amplitude object (Newport USAF resolution test target) and a subpixel cross-correlation image registration algorithm to find the displacements for every image. In this way, we can confirm that we are working in the linear regime of the motor as well as estimate the repeatability and accuracy of the procedure. This procedure only needs to be done once after setting up the interferometric arrangement.

## 2.3 SSC algorithm

The developed SSC algorithm for retrieving the complex amplitude distribution of the three FOVs involved in the process starts with the recording of four off-axis holograms (two in horizontal and two in vertical directions) in time sequence, where each hologram is further processed as follows: i) FFT of the hologram, ii) Fourier filtering of one of the diffraction orders to retrieve the cross-correlation of the complex fields involved in the hologram, iii) centering at the Fourier domain of the filtered complex spectrum, and iv) inverse FFT. Thus, each cross-correlation can be denoted as:

$$\hat{U}(u,v) = \hat{O}_1(u,v) \otimes \hat{O}_2^*(u,v), \quad (1)$$

where ($\otimes$) denotes convolution and $\hat{O}_1(u,v)$ and $\hat{O}_2(u,v)$ are the Fourier transforms of the first and second complex FOVs involved on each recorded hologram, respectively, and $u$ and $v$ are the spectral coordinates.

In conventional off-axis configuration, the complex field of one of the beams (let's say $\hat{O}_2(u,v)$) is known (usually a plane wave) making possible to recover the complex field coming from the other one (containing info of the sample). Our purpose is to use arbitrary and unknown complex fields on both beams and retrieve full QPI on both complex images. This is possible by using some of the properties of the complex Cepstrum, initially proposed by Oppenheim and Shafer [49] as solution to address a different deconvolution problem in signal processing. We will denote the complex Cepstrum $\tilde{G}(x)$ of a function $G(x)$ as:

$$\tilde{G}(x) = \mathcal{F}^{-1}\{\log(\mathcal{F}\{G(x)\})\} \quad (2)$$

where $\mathcal{F}$ and $\mathcal{F}^{-1}$ denote, respectively, Fourier and inverse Fourier transform operations and $log$ is the natural logarithm.

By applying Eq. (2) directly into Eq. (1) we get:

$$\tilde{U}(u,v) = \tilde{O}_1(u,v) + \tilde{O}_2^*(u,v) \quad (3)$$

thus, allowing access to the addition of the complex spectrums. However, our Cepstrum-based algorithm involves the recording of 2 off-axis holograms in the same direction, where the second hologram presents a lateral shift in one of the complex images (leaving static the other one). Thus, assuming horizontal displacement (x direction), we denote the complex field as $O_1(x,y)$ and its shifted version as $O_{1,x}(x,y) = O_1(x-x_0,y)$. For this shifted hologram, Eq. (3) can be rewritten as:

$$\tilde{U}_x(u,v) = \tilde{O}_1(u-u_0,v) + \tilde{O}_2^*(u,v) \quad (4)$$

being $\tilde{O}_1(u-u_0,v)$ the spectrum of the 1-pixel shifted image. After subtracting Eq. (4) from Eq. (3) and making use of the complex Cepstrum and Fourier transform properties, we get to:

$$\tilde{O}_1(u,v) = \frac{\tilde{U}(u,v) - \tilde{U}_x(u,v)}{\{1 - \exp(-i2\pi u x_0)\}} \tag{5}$$

We are looking to recover both complex fields, $O_1$ and $O_2$, and this is accomplished since the complex Cepstrum is reversible. Thus, expanding Eq. (5) and after some algebra, we can find the relationships for the amplitude and phase, Eqs. (6)-(7) respectively, of the complex field $O_1(x,y)$ in terms of $U(x,y)$, $U_x(x,y)$ and the shift $x_0$:

$$|O_1(x,y)| = \exp\left(F^{-1}\left\{Re\{\tilde{O}_1(u,v)\}\right\}\right)$$
$$= \exp\left(F^{-1}\left\{\frac{F\{\log(|U(x,y)|) - \log(|U_x(x,y)|)\}}{1 - \exp(-i2\pi u x_0)}\right\}\right) \tag{6}$$

$$\varphi_{O_1}(x,y) = \mathcal{F}^{-1}\left\{Im\{\tilde{O}_1(u,v)\}\right\} = \mathcal{F}^{-1}\left\{\frac{\mathcal{F}\{\varphi_U(x,y) - \varphi_{U_x}(x,y)\}}{1 - \exp(-i2\pi u x_0)}\right\} \tag{7}$$

where $\varphi_{O_1}$, $\varphi_U$ and $\varphi_{U_x}$ are the phases for the complex fields $O_1(x,y)$, $U(x,y)$ and $U_s(x,y)$, respectively.

Finally, $O_2(x,y)$ can be easily recovered once $O_1(x,y)$ is known by subtracting $\tilde{O}_1(u,v)$ from Eq. (3) and doing an inverse Fourier transform operation and $log^{-1}$ (or $exp$) in the resulting complex distribution $\tilde{O}_2^*(u,v)$. The same process is carried out considering the complex fields $O_2(x,y)$ and $O_3(x,y)$, where the shifted complex field is $O_3(x,y)$, to finally achieve the third FOV from the other two off-axis holograms recorded having a vertical shift of one pixel.

Equation (5) clearly identifies the zero-ambiguity problem. For those spatial frequencies where the denominator becomes 0, the retrieved object spectrum tends to infinity. This happens for all the positions where $ux_0$ is integer, which in the case $x_0 = 1$ is accomplished only when $u = 0$. Something similar will happen when consider the shift in the second multiplexing direction. These two lines of useless information are reduced to the central pixel (and its close neighborhood) at the Fourier domain by combining both multiplexed directions with the bow-tie masking procedure. This is what we called zero-ambiguity at the Fourier domain which finally provides background inhomogeneities when performing inverse FFT. However, if $x_0 > 1$ there will not be a single ambiguity line at $u = 0$ but other ones would appear at values of $u$ satisfying that the product of $ux_0$ results in an integer value. Even though SSC algorithm could still be applied taking in account those singularities, choosing $x_0 = 1$ minimizes the number of ambiguity lines, thus relaxing the constraints at the Fourier domain for applying Fourier filtering.

### 2.4 Samples used for validation of the method

The phase resolution test target used to show step-by-step the proposed methodology as well as to provide its QPI validation, see Results section, is from Benchmark Technologies [50]. Phase target features are incrementally taller in 50 nm nominal increments (as is depicted at beginning of Fig. 2) and estimated height values are directly obtained from atomic force microscopy (AFM) measurements on the phase target according to the information provided by the manufacturer. Static biosamples include LnCaP and PC3 prostate cancer cell lines and cheek cells. The prostate cancer cells were prepared according to the following procedure. The cells were cultured in RPMI 1640 medium with 10% fetal bovine serum, 100U/ml Penincillin and 0.1ug/ml Streptomycine at standard cell culture conditions (37°C in 5% CO2 in a humidified incubator). Once the cells reached a confluent stage, they were released from the culture support and centrifuged. The supernatant fluid was discarded by centrifugation and the cells were resuspended in a cytopreservative solution and mounted in a microscopy slide for its analysis. The cheek cells were directly extracted from the inner side of the cheek of one of the coauthors and were free donated to the experimental validation here provided. The cells were collected from the cheek by a cotton swab and smeared onto a microscope slide.

### 3. Results

### 3.1 Experimental validation with calibrated phase objects

Complementing the experimental results included in Fig. 2, Fig. 3 and Fig. 4 provide quantitative phase analysis of the three recovered FOVs by CIM. Fig. 3 (a.1) presents the static image (blue FOV in Fig. 2) in positive QPI mode while Fig. 3 (a.2) and Fig. 3(a.3) include the ROI (dashed black rectangle in a.1) in QPI and 3D thickness

(computed from retrieved phase values and considering the relation $\varphi(x,y) = \frac{2\pi}{\lambda} \cdot \Delta n \cdot t$, being $\lambda$ the illumination wavelength, $\Delta n = n_{test} - n_{air}$ the refractive index difference between the test elements and the air, and $t$ the thickness of those elements) visualization modes, respectively. To quantitatively compare the reconstructions by CIM, Fig. 3(b) shows the reconstruction of the same USAF test pattern obtained with a conventional off-axis DHM layout (a Mach-Zehnder configuration parallelly assembled in the lab using almost the same optical elements) as ground truth method for validation purposes.

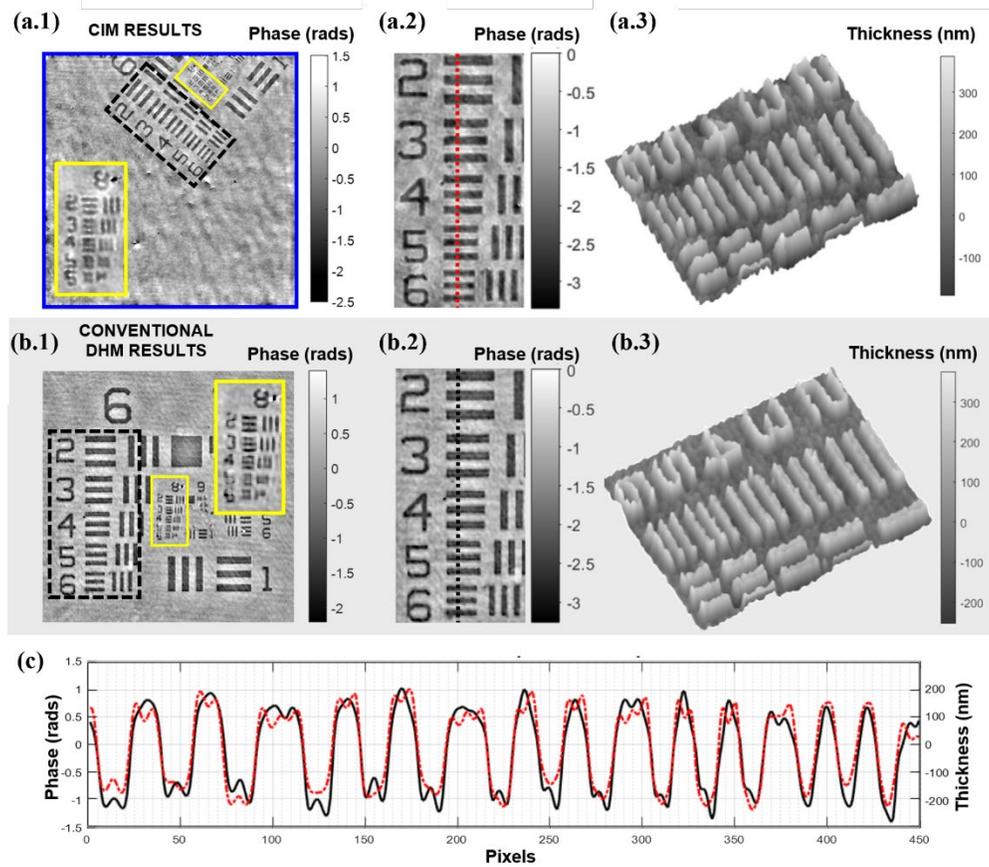

Fig. 3. Static image analysis. (a) Experimental images provided by CIM where (a.1) presents the full static image retrieved FOV in QPI mode, (a.2) includes a magnification of the ROI marked with a dashed black line rectangle in (a.1), and (a.3) shows the thickness profile of the defined ROI. (b) Same experimental analysis from the QPI provided by a conventional Mach-Zehnder off-axis DHM (ground truth method). (c) Comparative quantitative plot along the elements 2-6 of the group 6, (red dashed line for CIM, continuous black for ground truth). In addition, yellow rectangles at (a.1) and (a.2) show the last resolved elements (Group 8-Element 4) by both methods.

It is worthy to notice as resolution limit is preserved in the proposed CIM methodology. As the yellow insets at Figs. 3(a.1) and 3(b.1) depict, resolution limit is defined on both arrangements by Group 8-Element 4 corresponding with a spatial frequency of 362 lp/mm (or equivalently, 2.76 μm in spatial resolution). This value is in good agreement with the theoretical prediction ($\rho_{theo}$ = k λ/NA = 0.82 x 0.633 / 0.21 = 2.47 μm) since the following element (Group 8-Element 5) defines a resolution limit of 406 lp/mm (or 2.46 μm) which is not resolved in any of the two compared imaging modalities.

In addition to the visual comparison, Fig. 3 (c) includes the plots along the dotted vertical lines at Fig. 3 (a.2) and Fig. 3 (b.2) for the proposed method (red dashed plot) and the ground truth (continuous black plot). The plots compare efficiently and show good agreement between the two methods and with the thickness estimate provided by the test manufacturer. Table 1 shows the quantitative comparison where the thickness is estimated from the retrieved phase values (see previous introduced relation), the illumination wavelength and the refractive index of the polymer constituting the target. The agreement between these measurements shows the ability of the proposed method to retrieve accurate quantitative data. Moreover, the signal-to-noise ratio

(SNR) was also computed from the images included at Figs. 3(a.2) and 3(b.2) and the results are also included in Table 1.

Table 1. Quantitative analysis of CIM using the phase target. Comparison of CIM with conventional off-axis DHM for the static image (blue FOV).

|  | Thickness ± SD (nm) | SNR |
|---:|:---:|:---:|
| Nominal | 300 |  |
| Estimated by AFM | 329.5 |  |
| Measured by CIM | 316 ± 26 | 8.9 |
| Measured by conventional DHM | 332 ± 24 | 10.0 |

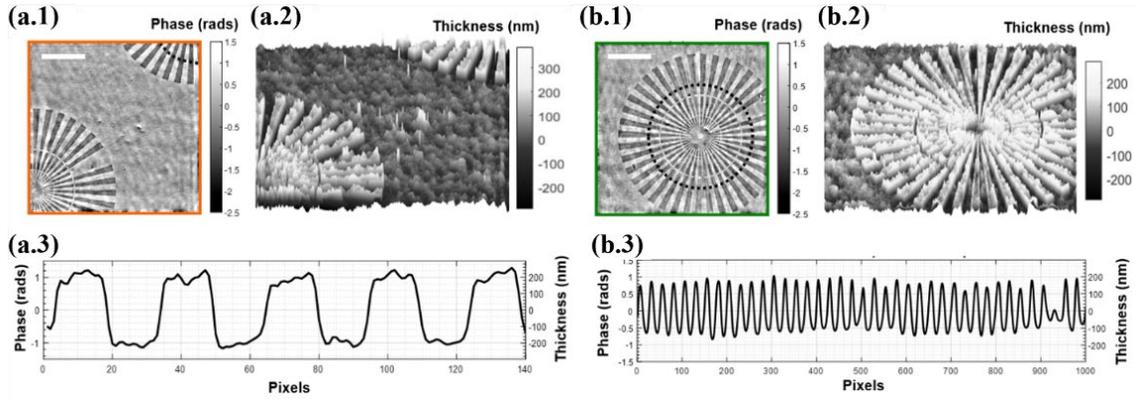

Fig. 4. Analysis of the additional FOVs retrieved by CIM. (a) Horizonal image (orange FOV) in (a.1) QPI and (a.2) 3D thickness visualization modes, and (a.3) plot along the dotted black line in the upper right target at (a.1). (b) Same analysis for the vertical image (green FOV). White scale bar at a.1 and b.1 are 100 microns length.

Since the proposed CIM methodology triples the FOV, similar analysis can be done with the additional complex images (orange and green FOVs). Thus, Fig. 4 presents the experimental results concerning QPI, 3D thicknesses view and thickness profile plot for both the horizontal (orange) and vertical (green) retrieved FOVs. For the horizontal image, we have computed the plot along the focus star pattern with 350/388 nm nominal/estimated heights (upper-right dotted line in Fig. 4 (a.1)) with a result of 400 ± 24 nm. And for the vertical case, the plot is done with a test pattern with 250/278 nm of nominal/estimated heights (dotted circular line at Fig. 4 (b.1)), obtaining 287 ± 24 nm. Again, these values are in good agreement with the estimated ones.

### 3.2 Experimental validation with biological samples

CIM has also been validated with static biosamples (two different prostate cancer cell lines – LnCaP and PC3 – and cheek cells). Fig. 5 presents the experimental results when considering the LnCaP cells. Starting from column a where two representatives of the four required FOVs (with internal magnification black rectangle ROI) in horizontal and vertical directions are included, CIM retrieves the complex field of the static image (blue FOV in column b) considering the set of four images (the two ones included in column a plus the pair of 1-pixel shifted complex images) and following the working scheme included in Fig. 2. After that, column b also includes the complementary complex FOVs in horizontal (orange frame) and vertical (green frame) directions, which are extracted by subtracting the initially retrieved FOV (blue frame) to each of the two complex fields from the orthogonal directions. These are the outputs of proposed CIM allowing QPI of the inspected sample while tripling the FOV.

These CIM results are now compared to conventional off-axis DHM as a ground-truth method for QPI. However, to assemble a parallel station based on conventional off-axis Mach-Zehnder or Michelson layouts would introduce some drawbacks from a comparison point of view. On one hand, the sample must be moved from one arrangement to another meaning that different focusing conditions will happen. Moreover, it will be difficult to localize the same cells for the comparison. And on the other hand, different components (beam splitters, lenses, etc.) are likely to be introduced, meaning that additional differences could be originated by the new optical elements and not from the compared techniques. For this reason, we have compared the results with off-axis common-path DHM, which is also a well-established method for QPI (see Refs. [11–17,32,33,37,38]) and has the advantages of using exactly the same optical components and assuring that the sample is focused on the same section. This type of off-axis common-path DHM can be easily implemented by keeping the same static FOV and changing the off-axis angle of the other interferometric beam so that a clear (without cells) area of the sample is used as the reference beam. That is, a surrounded free space on the sample acts as the reference beam for the off-axis interferometric recording. The drawback is that, even though there are no cells in the clear area acting as the reference beam, there may be some phase disturbances induced by the cell fixation process itself, and it cannot be treated as a pure clean reference beam as in conventional off-axis DHM with an external reference beam. However, this type of common-path arrangement has been widely validated in the past, and therefore it is justified to use it for direct comparison. Thus, column c shows the same FOVs retrieved after using common-path DHM in off-axis mode on the same experimental layout. As can be seen, results from common-path DHM are in good agreement with those retrieved by proposed CIM when comparing the phase values. However, CIM results appear to be slightly more noisy than common-path DHM ones. To compare the reconstructions, the difference between the outcomes from CIM and common-path DHM for the static FOV is presented in column d, showing a mean ± SD value of -0.02 ± 0.24 rads. This difference shows a very nice agreement between these two methods, where the standard deviation (SD) value means a 3.8% of error when considering a full $2\pi$ phase cycle.

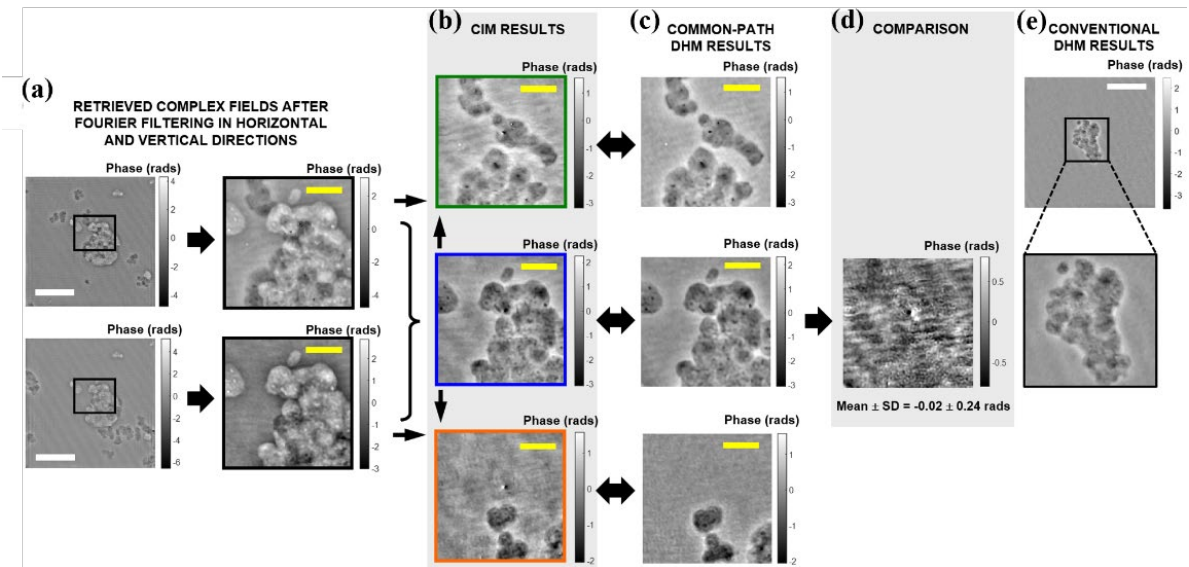

Fig. 5. QPI comparative for LnCaP cells using different DHM modalities. (a) Retrieved complex fields in horizontal and vertical directions from which CIM is applied including a magnified ROI where cell overlap occurs. (b) Retrieved results by CIM for static (blue), horizontal (orange) and vertical (green) complex FOVs. (c) Same ROIs retrieved by common-path DHM in off-axis configuration. (d) Comparison (subtraction) between the CIM and the common-path DHM recovered images for the static (blue) complex FOV. (e) Results provided by conventional off-axis DHM (Mach-Zehnder arrangement). White and yellow scale bars are 100 and 25 microns length, respectively.

In addition, the reconstruction of the same biosample coming from conventional off-axis DHM is included in column *e* of Fig. 5. Unfortunately, results do not include the same cell cluster (it was not possible to localize the same one over the whole slide) but a very similar one. According to the images in Fig. 5, the three methods provide very similar results both qualitatively and quantitatively, but the idea here is to quantify the phase noise obtained when using the three different reconstructions methods. For this, we have computed the mean value and its SD in some areas where no cells are present in the reconstructed images coming from CIM, common-path DHM and conventional DHM. The results are included in Table 2 showing good agreement with each other, with small differences that will be discussed later on.

Table 2. Quantitative analysis of CIM using static cells. Comparison of CIM with common-path and conventional off-axis DHM for some background areas without cells.

| Method | Mean (rads) | ± SD (rads) |
|---|---|---|
| CIM | 0.03 | 0.21 |
| Common-path DHM | 0.01 | 0.11 |
| Conventional DHM | 0.01 | 0.08 |

Unlike LnCaP cells, PC3 cells are individual cells showing no aggregates with other cells, thus adding a different casuistic to be studied. Similar to columns a-b-c at Fig. 5, Fig. 6 includes the main outcomes when considering PC3 cell line. Two (one in horizontal and other in vertical directions) of the four complex images needed to apply CIM are included in column a, where the ROI marked with a black rectangle shows the overlap of the same static cells for both images. The resulting phase images concerning full FOV, same ROI (now white rectangle), and 3D view of the cells after applying CIM are presented in column b. And to allow direct comparison, column c includes the common-path DHM results from the red rectangles included in the full FOV images in column a, where no cells are present in one of the interferometric beams. Again, CIM allows separation of the cells coming from the different FOVs and shows very similar results compared to common-path DHM.

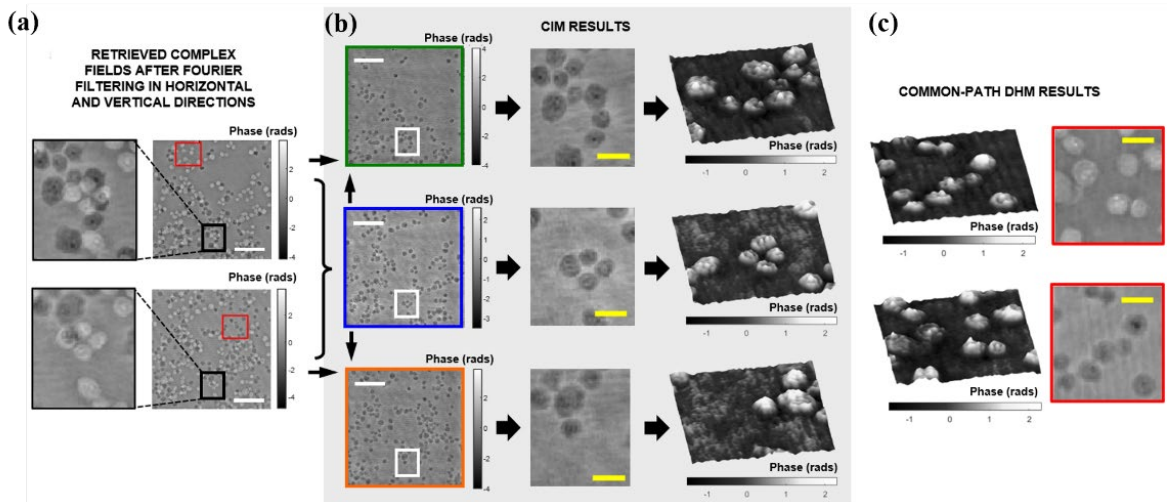

Fig. 6. QPI comparative for PC3 cells using different DHM modalities. (a) Retrieved complex fields in horizontal and vertical directions from which CIM is applied including a magnified ROI where cell overlapping are visible. (b) Retrieved results by CIM for the static (blue) and horizontal (orange) and vertical (green) complex FOVs including 3D view of the retrieved phase values. (c) Results considering different cells that are overlapping with clear areas (corresponding to the red rectangles in a) which can be processed using common-path DHM for direct comparison. White and yellow scale bars are 100 and 25 microns length, respectively.

Finally, cheek cells are also considered to validate the proposed CIM concept. As in previous biosample cases, Fig. 7 summarizes the main outcomes including two of the four complex images from which the three complex fields are retrieved. Once more, CIM enables perfect separation and QPI of the different complex fields involved in the process.

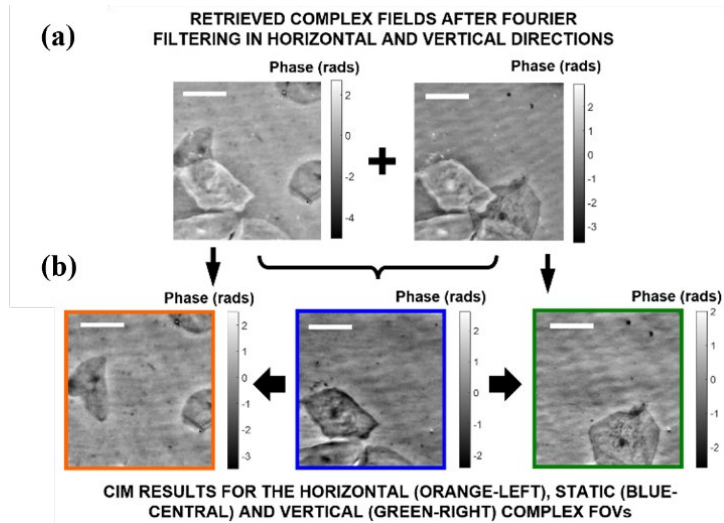

Fig. 7. Results coming from CIM for cheek cells. (a) Retrieved complex fields in horizontal and vertical directions from which CIM is applied. (b) Retrieved results by CIM for the static (blue) and horizontal (orange) and vertical (green) complex FOVs. White scale bars are 50 microns length

## 4. Discussion

This paper presents an innovative procedure for retrieving QPI breaking the general conception in most of the fields of interferometry of having one of the interferometric beams as a clear reference beam with no information. Thus, no restrictions are introduced in the interferometric beams regarding sample constraints (to be sparse, containing specific blank areas, etc.) and no bulky/complex arrangements are needed to synthesize a clear reference beam. Due to this, the proposed methodology (named as CIM) is a potential candidate to be adapted to existing commercial bright-field microscopes for updating them with coherence sensing capabilities introducing minimal modifications, which is a current trend for democratizing science by allowing QPI in biological experiments to those laboratories with constrained budgets. Although being presented for QPI in DHM, the proposed methodology is potentially adaptable to other coherent imaging disciplines such as digital holography, optical coherence tomography, lensless microscopy or tomographic phase microscopy, just to cite a few, where interference between two beams is inherent, thus opening new future perspectives of applicability.

The proposed methodology increases flexibility in the recording interferometric process by avoiding the reentering of an external reference wave for the interferometric recording. This is a significant improvement regarding i) common-path interferometric layouts, where sample constraints (sparse, confined, etc.) must be imposed to transmit in parallel the reference beam with the imaging beam through the imaging system [11-17], since interference is performed considering two arbitrary complex fields of the sample, and ii) the external introduction of the reference beam as in classical Mach-Zehnder arrangements since mechanical and/or thermal instabilities are not affecting the process, thus opening the door to use partially coherent light sources.

CIM is based on the acquisition of a coherent set of interferograms (4 holograms in the presented validation) from which the complex field of the cross-correlation term between complex amplitudes is firstly retrieved. We have used here an off-axis configuration to simplify this part but any other interferometric arrangement (on-axis, slightly off-axis, etc.) can be used as well. The use of off-axis recording involves Fourier domain filtering where the width of the circular mask doubles the one used in classical off-axis DHM, since we are retrieving the cross-correlation of the two complex fields involved in the interferometric recording. This fact constraints the pixel size of the digital sensor since empty resolution is needed at the Fourier domain to filter out the desired spectral distribution without including zero order term disturbances. In that sense, the proposed methodology restricts the space-bandwidth product of the system in a similar way than off-axis DHM arrangements also do. Additional multiplexing comes from the temporal degree of freedom because of the need to record 4 images with static sample. As consequence, the FOV is improved (a factor of 3) as well as complex values are retrieved.

In that sense, modern QPI methods have appeared to optimize spatial bandwidth usage. Some examples are Hilbert-Huang approaches [15, 29] and Kramers-Kronig relations [48-49]. While Hilbert-Huang methodology is capable of working in a single-shot, it needs the introduction of an external reference beam. However, Kramers-Kronig approaches are non-interferometric and non-iterative imaging methods for retrieving complex values from two intensity images based on aperture shaping using either a spatial light modulator in the Fourier plane of the system [48] or by illumination scanning (oblique illumination) using galvanometric mirrors and a number of intensity images ranging from 2 to 6 depending on the target application and inspected sample [49]. A novel concept aimed to complex amplitude reconstruction has been recently presented involving a coupled configuration for combining two in-line holograms and one off-axis hologram with a rapidly converging iterative procedure based on two-plane coupled phase retrieval method [50]. The main advantage is that no constraint or prior knowledge of the object is needed (including support, non-negative, sparse constraints, etc.) and the algorithm needs two in-line holograms at two different distances plus one off-axis hologram at one of the two previous distances. From these three holograms, it is possible to retrieve QPI but each one of the holograms need the introduction of an external reference beam. Again, our proposed methodology needs a comparable number of images as in Refs. [48-50] while our algorithm is completely innovative.

The presented procedure requires 2+2 off-axis holograms in orthogonal directions where, for each direction, the two images are one arbitrary and the other must introduce a 1-pixel camera shift in one of the interferometric images, leaving the second one static. In the proposed implementation, we have considered horizontal and vertical directions but any other pair of orthogonal directions can be used or even non-orthogonal directions can also be considered. The need for a second direction comes from the presence of an ambiguity line with useless information in the retrieved Fourier spectrum for each direction. Thus, the use of two directions (orthogonal or non-orthogonal) reduces the ambiguity problem to the DC term. This zero-ambiguity problem is similar to the one provided by TIE algorithms [48] providing a smeared in the background of the retrieved phase image (see step 8 at Fig. 2). Although other strategies can be followed (Zernike functions, polynomial fitting, machine learning, etc.), we have adopted high-pass filtering process due to its simplicity. Here, we have implemented a gaussian filter with a FWHM equal to 9 pixels. This value is appropriate for background homogenization without affecting the low spatial-frequency content of the samples as comparison plot at Fig. 3(c) verifies. Note that the rendering of the bars elements of Group 6 is quite similar when comparing CIM and conventional off-axis DHM layout and, if low spatial-frequency content of the object would be lost, the bars in the reconstructed image provided by CIM would be empty showing a similar value than the background.

After retrieving the complex amplitude distribution of the cross-correlation term for the 4-hologram set by using conventional Fourier domain filtering, a novel algorithm (named as SSC) based on Cepstrum subtraction in combination with a complementary Fourier domain bow-tie masking, allows the complex pupil recovery of the static image. After that, the other two additional images involved in the process are easily recovered by subtraction, thus tripling the FOV provided by the conventional imaging system. Although the involved FOVs are not adjacent in our proposed CIM implementation, this fact could be accomplished by properly adjusting the off-axis angles and the distance from the off-axis mirror to the sensor. Even though, a non-contiguous FOV extension can be still useful in applications where processing of images of large areas is required (as in tomographic flow cytometry [51], for instance) because it allows to cover a larger area in a single interferogram despite being non-contiguous.

Cepstrum-based operations have been widely employed, first for speech component deconvolution [49] and later for several applications in signal processing such as seismology, hydroacoustic, echo detection and noise cancellation [52–55], as well as in image processing [56,57], just to cite a few. Although its application to DHM has been also reported [45–47], those proposed cepstrum-based algorithms apply only for the removal of unwanted diffraction orders in the Fourier domain of the recorded hologram in order to relax the constraints regarding space bandwidth product in the system. In our proposed methodology, we are dealing with a completely different problem since we are decoupling the complex amplitude distributions of the two interferometric beams being both of them arbitrary (no need to introduce a clean reference beam for the recording process). This fact not only allows both interferometric beams to carry useful complex information of the sample, increasing the effective FOV, but also opens entirely new possibilities on common and quasi-common path approaches in DHM.

The proposed CIM proof-of-concept validation has been implemented using a compact quasi-common path Michelson interferometric configuration assembled at the output port of a classical infinity corrected microscope configuration mounted on an optical table. In the interferometric layout, one of the mirrors is movable to change the off-axis direction (needed for the two orthogonal directions to be processed) as well as to allow the pixel shift needed for the application of the SSC algorithm. On one hand, we have selected a Michelson interferometric arrangement due to its compactness but any other interferometric architecture (Mach-Zehnder, common-path, etc.) could be also implemented depending on system requirements (tube lens focal length, disposable device to allow 1-pixel shift, etc.). On the other hand, the change in the off-axis orthogonal direction as well as the 1-pixel shift can be performed by, for instance, a motorized gimbal mount in a very robust way. However, in this experimental validation, we have decided to implement a tiltable mirror mount which is manually switched between the two orthogonal directions in combination with a linear translation stage to provide a 1-pixel shift in one of the interferometric images for each direction. This 1-pixel shift needs to be accurately done and, for this, we have implemented a calibration procedure where a continuous recording of images while simultaneously shifting the image allows the precise selection of the proper axial displacement in the translation stage producing the 1-pixel lateral shift required for the method. This process only needs to be done once as preliminary system calibration, so the recording of the 1-pixel shifted images can be done very fast.

In terms of temporal resolution, the number of required holograms (4 in total) is similar to the number of holograms used in phase-shifting procedures for complex amplitude retrieval in on-axis DHM. Moreover, the equipment needed to provide the 1-pixel shift in one of the images is similar to the devices needed in phase-shifting implementations. Here a linear translation stage (previously implemented in phase-shifting strategies – see for instance Ref.[14]) has been used, since axial displacement in one of the mirrors (M2 in Fig. 1) allows lateral shift due to the off-axis geometry. Thus, the proposed CIM methodology is at the same level than widely- and commonly-used phase-shifting strategy as long as the switch between orthogonal directions could also be quickly done (for instance, using motorized devices) in terms of temporal resolution but tripling the FOV in comparison with phase-shifting methods.

The reported experimental validation of CIM used static samples involving an object (phase resolution test target) with topographic information to quantitatively validate the results and biological samples (two lines of prostate cancer cells and cheek cells) to demonstrate the performance of the proposed method in comparison with well-established classical DHM platforms (common-path and conventional Mach-Zehnder DHM). The obtained results reasonably compare with classical DHM methodologies although SNR and SD values (see Table 1 and Fig. 5(f)) are slightly worse than those retrieved by classical DHM. Strictly speaking, the comparison with common-path DHM is the most appropriate because both methods share the same optical layout (optical elements, focusing section at the sample plane, etc.). However, since common-path DHM uses a clear region of the sample acting as reference beam, that area can introduce additional phase disturbances in the reconstruction. For this reason, classical Mach-Zehnder DHM is also included in this comparison but, in this case, the layout is not exactly the same one: although almost the same optical components (objective, tube lens, beam splitter and mirrors) are used, additional elements (identical microscope lens, beam splitter, mirror and tube lens) are needed. Moreover, it is impossible to ensure the focusing at the same sample section since the sample is moved from one arrangement to another. Whatever the case, common-path and classical DHM provide similar results while CIM reports slightly worse values. We believe that the zero-ambiguity problem can be one of the causes for this discrepancy as it can be seen as differences in the background areas of the retrieved images (see for instance, Fig. 5 columns (b)-(c)). But it is not the reason behind the high frequency mismatch also visible at those figures. Maybe the discrepancy in high frequency content comes from the violation of the basic principle of CIM caused by coherent noise. Since coherent noise is not constant between the set of interferograms needed to extract the complex amplitude distribution of the static image, there are not the same noise conditions when implementing the 1-pixel shift in one of the FOVs for applying the SSC algorithm (steps num. 6-7). This minimum difference could yield in the generation of some unwanted discrepancies when extracting the static image that will be translated into the additional FOVs when performing subtracting operation (steps num. 10-11). In any case, those degradations seem to be marginal and do not prevent the use of CIM for most of its potential in QPI applications.

## 5. Conclusions

In summary, we have reported on a new methodology for interferometry-based QPI where no restrictions about the interferometric beams are required, that is, without a priori assumptions (clear beam, previously known complex distribution, etc.) about the reference beam. Thus, any pair of coherent fields can be used in the interferometric process, including a shifted version of the target sample itself. The proposed CIM uses four interferograms to retrieve the complex amplitude distribution of the three FOVs involved in the process, so besides of retrieving the QPI of the "imaging beam" (following a classical DHM nomenclature), it also triples the useful FOV since it retrieves the two additional "reference beam" FOVs involved in the process. The novel numerical procedure for the tripled-FOV QPI implements a phase-shifting Cepstrum-based (SSC) algorithm where a total of four off-axis holograms are the input of a normalized Cepstrum subtraction operation yielding in the recovery of the static complex field, from which the additional FOVs are afterwards retrieved. Experimental validation of this new QPI methodology is qualitatively and quantitatively reported using different types of samples and the results satisfactorily compare with conventional DHM. Next steps will be directed to explore the proposed method with a different range of microscope objectives (different magnification and NA values) and different interferometric configurations, as well as to improve the technique towards fast imaging (video rate) implemented in real microscope embodiments.

**Authors' contributions**

**Ricardo Rubio-Oliver**: Formal analysis; Methodology; Software; Experimental validation; Writing – original draft. **Javier Garcia**: Conceptualization; Data analysis; Validation; Supervision; Writing – review & editing. **Zeev Zalevsky**.: Conceptualization; Validation; Writing – review & editing. **José Ángel Picazo-Bueno**.: Investigation; Data curation; Validation; Software; Writing – review & editing. **Vicente Micó**: Methodology; Supervision; Validation; Visualization; Project administration; Resources; Funding acquisition; Writing – original draft; Writing – review & editing.

**Funding.** This research has been funded by the Grant PID2020-120056GB-C21 funded by MCIN/AEI/10.13039/501100011033. J. A. Picazo-Bueno is supported by the Spanish grant "Margarita Salas" (Ref. MS21-100), proposed by the Ministry of Universities of the Government of Spain (UP2021-044).

**Disclosures.** The authors declare no competing financial interest.

**Data availability.** The datasets used and analyzed during the current study are available from the corresponding author on reasonable request.